\documentclass[10pt,twocolumn,letterpaper]{article}

\usepackage{iccv}              
\usepackage{graphicx}
\usepackage{multirow}
\usepackage{booktabs}
\usepackage{diagbox} 
\usepackage[accsupp]{axessibility}
\usepackage{amssymb}
\usepackage{colortbl}

\definecolor{iccvblue}{rgb}{0.21,0.49,0.74}
\usepackage[pagebackref,breaklinks,colorlinks,allcolors=iccvblue]{hyperref}
\usepackage{bbding}
\usepackage{pifont}
\usepackage{fontawesome5}
\title{All That Glitters Is Not Gold:\\Key-Secured 3D Secrets within 3D Gaussian Splatting}

\author{
{Yan Ren}\textsuperscript{1} \quad {Shilin Lu}\textsuperscript{1} \quad {Adams Wai-Kin Kong}\textsuperscript{1}\\
\textsuperscript{1}Nanyang Technological University, Singapore \\ 
\tt\small allwillhaveof@gmail.com,
\tt\small shilin002@e.ntu.edu.sg, \tt\small adamskong@ntu.edu.sg\\
}

\begin{document}
\maketitle
\begin{abstract}
Recent advances in 3D Gaussian Splatting (3DGS) have revolutionized scene reconstruction, opening new possibilities for 3D steganography by hiding 3D secrets within 3D covers. The key challenge in steganography is ensuring imperceptibility while maintaining high-fidelity reconstruction. However, existing methods often suffer from detectability risks and utilize only suboptimal 3DGS features, limiting their full potential.
We propose a novel end-to-end key-secured 3D steganography framework (KeySS) that jointly optimizes a 3DGS model and a key-secured decoder for secret reconstruction. Our approach reveals that Gaussian features contribute unequally to secret hiding. The framework incorporates a key-controllable mechanism enabling multi-secret hiding and unauthorized access prevention, while systematically exploring optimal feature update to balance fidelity and security. To rigorously evaluate steganographic imperceptibility beyond conventional 2D metrics, we introduce 3D-Sinkhorn distance analysis, which quantifies distributional differences between original and steganographic Gaussian parameters in the representation space.
Extensive experiments demonstrate that our method achieves state-of-the-art performance in both cover and secret reconstruction while maintaining high security levels, advancing the field of 3D steganography. Code is available at \href{https://github.com/RY-Paper/KeySS}{KeySS}.
\end{abstract}
\begin{figure}[!ht]
\centering
\small
\includegraphics[width=1.0\linewidth]{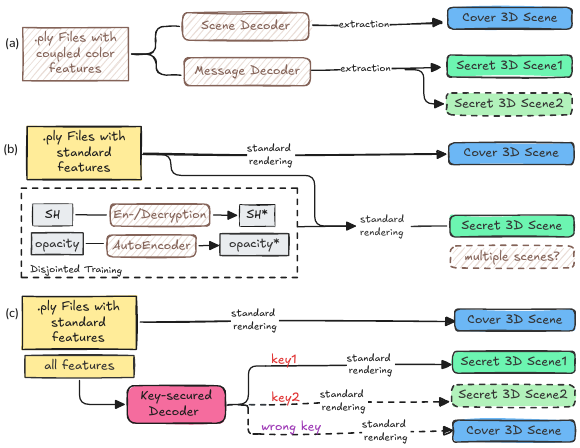}
\vspace{-0.4cm} 
\caption{Compared to existing methods like (a) GS-Hider~\cite{gshider_zhang2024gshider} and (b) WaterGS~\cite{sis_guo2024splats}, (c) the proposed method maintains the standard 3DGS format compatibility while achieving superior performance through fully exploiting inherent features for fidelity and implementing a key-controllable scheme that enables both multi-secret hiding and defense against incorrect key inputs.
}
\vspace{-0.2cm} 
\label{fig_method_comparisons}
\end{figure}
    
\begin{figure*}[!ht]
\vspace{-0.6cm} 
\centering
\small
\includegraphics[width=1.0\linewidth]{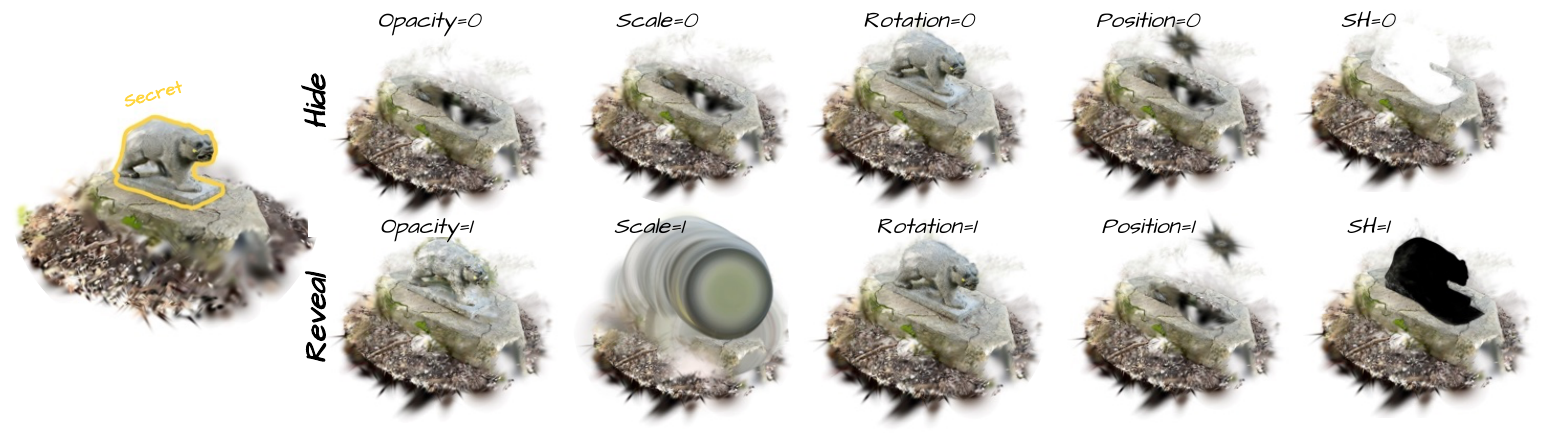}
\vspace{-0.8cm} 
\caption{
3D Gaussians provide rich steganographic potential through multiple attributes: opacity, scale, rotation, position, and spherical harmonics (SH). However, naive approaches that simply zero out specific attributes to hide secrets pose fundamental security risks. The presence of hidden content can be easily detected by simply restoring the zero-value attribute, instantly revealing the hidden content. This limitation motivates our exploration of optimal feature transformation strategies for both effective hiding and security.
}
\vspace{-0.2cm} 
\label{fig_motivation}
\end{figure*}
\section{Introduction}
\label{sec:intro}
Steganography~\cite{survey_classic_stegan_cheddad2010digital,stegan_hu2024learning,stegan_yu2023cross} constitutes a security methodology that conceals secret information within seemingly innocuous carriers such as images~\cite{survey_image_stegan_kadhim2019comprehensive,imgstegan_hamid2012image,imgstegan_subramanian2021image}, text~\cite{survey_text_stegan_majeed2021review,txtstegan_delina2008information,txtstegan_wu2024generative}, audio~\cite{survey_audio_dutta2020overview,audiostegan_djebbar2011view,audiostegan_djebbar2012comparative} and videos~\cite{survey_video_stegan_kunhoth2023video,video_in_video_mou2023large,videostegan_liu2019video,videostegan_sadek2015video}, which has demonstrated widespread applications in copyright protection~\cite{application_copyright_protection_megias2021data}, secure digital communications~\cite{application_digital_communication_varghese2023detailed}, and e-commerce systems~\cite{application_ecommerce_kumbhakar2023optimal}. 
The rapid advancements in 3D reconstruction technologies, such as nerual radiance fields (NeRF)~\cite{survey_nerf_gao2022nerf,nerfintro_bian2023nope,nerfintro_metzer2023latent} and 3DGS~\cite{survey_3dgs_fei20243d,3dgsintro_qin2024langsplat,3dgsintro_yu2024mip}, have catalyzed the development of 3D steganography, which has emerged as a promising paradigm for safeguarding 3D digital assets~\cite{survey_3Dstegan_girdhar2018comprehensive,3dstegan_zhang2023chaotic,3dstegan_zhou2021three,3dstegan_zhu2021gaussian}. 
Analogous to conventional steganographic techniques, 3D steganography aims to ensure that both the existence and content of secret messages remain imperceptible to unauthorized observers while maintaining reliable recovery of the concealed information. 

Despite advancements in 3D steganography with the emergence of 3DGS, existing 3DGS-based methods still face substantial limitations in practical applications.
GS-Hider~\cite{gshider_zhang2024gshider} modifies the standard 3DGS pipeline for the secret embedding: the coupled color features are utilized to replace the standard spherical harmonics (SH) coefficients and a scene decoder is introduced to replace standard rendering, introducing deviations from the standard GS pipeline (\cref{fig_method_comparisons}(a)).
While this approach achieves high fidelity for both cover and secret scenes, these modifications introduce noticeable artifacts that may raise suspicion among unauthorized users, ultimately compromising the system’s imperceptibility and security.
WaterGS~\cite{sis_guo2024splats} enhances imperceptibility through importance-graded SH encryption and autoencoder-assisted opacity mapping.
However, the separate concealment of SH coefficients and opacity results in a disjointed, non-end-to-end pipeline, limiting practical deployment. Additionally, the complexity of its secret embedding process prevents the encoding of multiple secret scenes within a single cover scene, reducing its flexibility and practicality in real-world scenarios.

To overcome these challenges, we introduce \textbf{Key}-\textbf{S}ecured 3D \textbf{S}teganography (KeySS), a novel framework that directly transforms cover 3D Gaussians to secret 3D Gaussians while preserving standard feature formats and rendering processes. Our approach integrates seamlessly with existing 3DGS pipelines while providing robust security through a key-controlled mechanism without compromising visual fidelity.
Our comprehensive analysis reveals a critical insight: \textbf{Gaussian features contribute unequally to steganographic effectiveness$\textendash$all that glitters is not gold.} As demonstrated in~\cref{fig_motivation}, opacity modifications effectively enable secret hiding while SH coefficients produce minimal impact or even destabilize the embedding process. However, relying solely on opacity creates a significant security vulnerability, as hidden information can be easily exposed by simply detecting and restoring zero-valued opacity attributes, compromising the entire steganographic system (\cref{fig_motivation}).
Based on these findings, we systematically explore optimal feature combinations that strategically balance reconstruction quality and steganographic imperceptibility (\cref{tab_ablation_sinkhorn}). This exploration identifies that combining Gaussian attributes of opacity with rotation, position and scale significantly improves both security and fidelity over single-feature approaches (\cref{fig_1hid1vis}).

To quantitatively evaluate security beyond conventional 2D metrics, we propose a 3D-Sinkhorn security evaluation metric, which analyzes distributional differences in the Gaussian parameter space itself. Our security analysis (\cref{tab_ablation_sinkhorn,fig_lowopacity}) confirms that while the opacity-only approach achieves reasonable rendering fidelity, they create distinctive statistical signatures that are 
invisible to conventional 2D metrics but 
detectable through our proposed distribution analysis, as shown in~\cref{fig_stegExpose}.
To summarize, KeySS makes the following contributions:
\vspace{0.2cm}
\begin{itemize}
\setlength{\itemsep}{3pt}
\setlength{\parsep}{0pt}
\setlength{\parskip}{0pt}
\item \textbf{End-to-End 3D Steganography Framework}: We introduce an end-to-end learning framework that jointly learns cover 3D Gaussians and a key-secured decoder for 3D secret hiding, while maintaining compatibility with standard 3DGS format and rendering pipeline.
\item \textbf{Key-Secured Decoder}: Our framework incorporates a key-controllable scheme that enables high-fidelity multi-secret recovery while ensuring security against unauthorized access.
\item \textbf{Fidelity-Security Balance Analysis}: We are the first to conduct systematic exploration of optimal 3D Gaussian feature combinations and introduce 3D-Sinkhorn distance as a novel security evaluation metric to balance fidelity and steganographic imperceptibility.
\item \textbf{Extensive Experimental Validation}: Experimental results demonstrate that KeySS achieves superior performance in terms of visual quality, reconstruction fidelity, and robustness against unauthorized extraction attempts.
\end{itemize}
\begin{figure*}[t!]
\centering
\small
\vspace{-0.6cm} 
\includegraphics[width=1.0\linewidth]{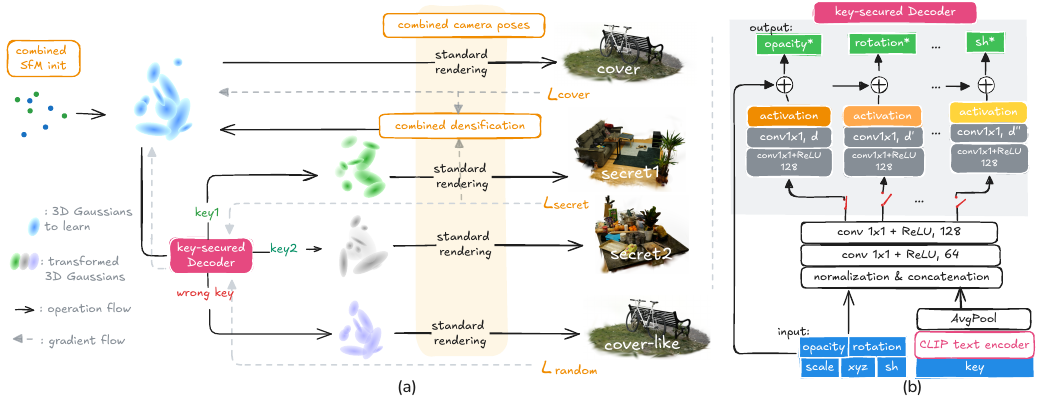}
\vspace{-0.6cm} 
\caption{(a) Our end-to-end 3D steganography framework jointly trains the cover 3D Gaussians and the key-secured decoder from scratch. To enhance training, we introduce combined camera poses for diverse training samples, combined SfM points for optimal initialization, and combined densifications for refinement.
(b) The key-secured decoder features a decoupled architecture with feature-specific layers for different Gaussian attributes. A key-controlled scheme enables multi-secret hiding and strengthens defenses against unauthorized extraction. Additionally, the feature-specific layers allow systematic exploration of the optimal feature update for secret embedding.}
\label{fig_flowchart}
\vspace{-0.2cm} 
\end{figure*}

\section{Background}
We briefly describe essential backgrounds. The rest of related work is deferred to~\cref{sec_sup_related_work}.
\subsection{3D Steganography}
3D steganography, using 3D models to hide secret messages, has been studied for decades~\cite{3dstegan_zhang2023chaotic,3dstegan_zhou2021three,3dstegan_zhu2021gaussian,survey_3Dstegan_girdhar2018comprehensive}. Traditional methods primarily modify 3D mesh geometry or topology for copyright protection, such as adjacent bin mapping~\cite{adjacent_bin_mapping_wu2009steganography}, triangle mesh reformation~\cite{triangle_reformation_thiyagarajan2013pattern}, and vertex decimation~\cite{vertex_decimation_tsai2014adaptive}. Other approaches retrieve hidden messages from 2D renderings of 3D distortions~\cite{3dto2d_yoo2022deep}. Recent advances in 3D reconstruction have shifted focus to more powerful representations, like NeRF and 3DGS, enabling new possibilities for steganography in neural rendering. For example, CopyNeRF~\cite{copy_nerf_luo2023copyrnerf} embeds copyright protection into NeRF models using watermarked color representations and a resistant rendering scheme, while WateRF~\cite{waterf_jang2024waterf} applies discrete wavelet transformation to both implicit and explicit NeRF models. NeRFProtector~\cite{nerfprotector_song2024protecting} introduces a plug-and-play strategy for protecting NeRF copyrights. However, these methods are primarily focused on embedding binary bit messages \cite{shilin_lu2025robust} with limited hiding capacity. StegaNeRF~\cite{stegan_nerf_li2023steganerf} pioneers higher-capacity hiding, including images and audio, within NeRF rendering. More recently, GS-Hider~\cite{gshider_zhang2024gshider} and WaterGS~\cite{sis_guo2024splats} have explored hiding 3D content in 3D reconstruction models. GS-Hider replaces 3DGS’s SH coefficients with coupled secure features and introduces separate decoders for the cover and secret scenes. However, these non-standard features and rendering process risk arousing suspicion and violating the imperceptibility principle of steganography. WaterGS, aligned with standard 3DGS rendering, uses importance-graded SH coefficient encryption and opacity mapping for secret embedding. However, by focusing primarily on SH and opacity features, it underutilizes the full 3DGS features and lacks end-to-end training, limiting flexibility in embedding multiple secrets in a single cover. Our proposed method, KeySS, aims to develop an end-to-end learnable 3D steganography framework that maintains both imperceptibility and flexibility.

\subsection{Preliminaries on 3DGS}\label{sec_method_preliminary}
Built upon the splatting technique, 3DGS~\cite{3dgs_kerbl20233d} models 3D scenes using a set of $\mathcal{N}$ anisotropic Gaussians: $\mathcal{G}=\{G_i(x)\}_{i=1}^{\mathcal{N}}$. These Gaussians are learned to capture the scene's structure and appearance, with attributes including center position $\boldsymbol{\mu}\in\mathbb{R}^3$, opacity $\alpha\in\mathbb{R}^1$, rotation $\mathbf{r}\in\mathbb{R}^4$, scale $\mathbf{s}\in\mathbb{R}^3$ and color $\mathbf{c}\in\mathbb{R}^{16\times3}$. Specifically, $\boldsymbol{\mu}$, $\mathbf{r}$, and $\mathbf{s}$ together describe the configuration of the $i$-th Gaussian:
\begin{equation}
G_i(\mathbf{x})=\exp\left(-\frac{1}{2}(\mathbf{x}-\boldsymbol{\mu}_i)^\top\boldsymbol{\Sigma}_i^{-1}(\mathbf{x}-\boldsymbol{\mu}_i)\right),
\end{equation}
where $\boldsymbol{\Sigma}_i=\mathbf{R}_i\mathbf{S}_i\mathbf{S}_i^\top\mathbf{R}_i^\top$ is the 3D covariance matrix defined by the scaling matrix $\mathbf{S}_i$ and rotation matrix $\mathbf{R}_i$.
Opacity $\alpha$ controls the transparency level of each Gaussian, and color $\mathbf{c}$ represents spherical harmonics (SH) to capture view-dependent appearance. For the standard rendering process from 3D to 2D, the 2D covariance matrix $\boldsymbol{\Sigma}_i'$ is formulated as $\boldsymbol{\Sigma}_i'=\mathbf{J}_i\mathbf{W}_i\boldsymbol{\Sigma}_i\mathbf{W}^\top_i\mathbf{J}^\top_i$, with the given viewing transformation matrix $\mathbf{W}$ and the Jacobian $\mathbf{J}_i$ of the affine approximation of the projective transformation. The final color of a pixel is calculated via alpha compositing:
\begin{equation}
    C=\sum_{i\in\mathcal{N}}{c_i}{\alpha_i'}\prod_{j=1}^{i-1}\left(1-\alpha_j'\right),
\end{equation}
where $\alpha_i'=\alpha_i\cdot\exp\left(-\frac{1}{2}(\mathbf{x}'-\boldsymbol{\mu}_i')^\top{\boldsymbol{\Sigma}_i'}^{-1}(\mathbf{x}'-\boldsymbol{\mu}_i')\right)$ represents the final opacity based on the projected coordinates $\mathbf{x}'$ and $\boldsymbol{\mu}'$. Like NeRF, 3DGS initializes with SfM points and employs adaptive density control (cloning/splitting/pruning) to enhance scene detail capture.

\section{Method}
\subsection{Problem Formulation}
We develop an end-to-end steganographic framework (\cref{fig_flowchart}(a)) that learns the transformation between cover and secret 3D Gaussians. More precisely, given $S+1$ sets of ground truth 2D images $\{I_\text{gt\_cover}^i\}_{i=1}^{\mathcal{M}_{cover}}$ and $\{\{I_\text{gt\_secret}^{i,s}\}_{i=1}^{\mathcal{M}_{secret}}\}_{s=1}^{\mathcal{S}}$ with aligned camera poses, our goal is twofold: (1) learn a 3DGS model $\mathcal{G}_\text{cover} = \{G_i(x)\}_{i=1}^\mathcal{N}$ that reconstructs the cover scene, and (2) learn transformations to decode $\mathcal{G}_\text{cover}$ into multiple 3DGS models $\mathcal{G}_\text{secret}^s = \{G_j^*(x)\}_{j=1}^\mathcal{N}$ that render as the secret scenes. This transformation is parameterized by a decoder $D$, such that:
\begin{equation}
\mathcal{G}_\text{secret}^s = D(\mathcal{G}_\text{cover}).
\end{equation}
To achieve both reconstruction fidelity and steganographic security, we jointly optimize $\mathcal{G}_\text{cover}$ and $D$ from scratch. This optimization is guided by a set of loss functions that enforce accurate reconstruction for both the predicted cover scene and the recovered secret scenes:
\begin{gather}
    \begin{split}
    \mathcal{L}_\text{cover}=&(1-\lambda_\text{cover})\mathcal{L}_1(I_\text{pred\_cover},I_\text{gt\_cover})\\
    &+\lambda_\text{cover}\mathcal{L}_\text{SSIM}\left(I_\text{pred\_cover},I_\text{gt\_cover}\right),
    \end{split} \label{eq:loss_cover} \\
    \begin{split}
    \mathcal{L}_\text{secret}^s=&(1-\lambda_\text{secret})\mathcal{L}_1(I_\text{pred\_secret}^s,I_\text{gt\_secret}^s)\\
    &+\lambda_\text{secret}\mathcal{L}_\text{SSIM}\left(I_\text{pred\_secret}^s,I_\text{gt\_secret}^s\right),
    \end{split} \label{eq:loss_secret}
\end{gather}
where $I_{\text{pred\_cover}}$ and $I_{\text{pred\_secret}}^s$ are rendered from $\mathcal{G}_{\text{cover}}$ and $\mathcal{G}_{\text{secret}}^s$ respectively. $\lambda_{\text{cover}}$ and $\lambda_{\text{secret}}$ are the trade-off coefficients between $L_1$ and SSIM losses. 

\subsection{Key-Secured Decoding Architecture}
The overview of our decoding architecture is shown in~\cref{fig_flowchart}(b). Specifically, our decoder is designed to build the transformation between the 3D Gaussians of the cover and secret scenes while maintaining a balance between fidelity and security.
To further enhance security, the decoder is conditioned on a user-specific key $k^s$,
ensuring that only authorized users with the correct key can accurately reconstruct the secret scenes. This can be formulated as:
\begin{equation}
\mathcal{G}_{\text{secret}}^s = D(\mathcal{G}_{\text{cover}}, k^s).
\label{eq:decoder}
\end{equation}

\noindent\textbf{The Key-Controllable Scheme} enables both multi-secret hiding capabilities and robust defense against unauthorized extraction attempts. We leverage CLIP's text encoder~\cite{clip_radford2021learning} to encode the keys, which excels at processing diverse textual inputs into semantic embeddings. The user-specific key $k$ is first tokenized and then processed through a transformer-based encoder $\mathbf{E}$ followed by average pooling operations to obtain the final key embedding: $\mathbf{k} = \text{AvgPool}(\mathbf{E}(k))$. The key embedding is concatenated with the normalized 3D Gaussian features as input to the decoder. To mitigate the risk of incorrect key attacks, the training process incorporates two key scenarios: (1) correct keys for secret recovery as defined in~\cref{eq:decoder}, and (2) randomly generated incorrect keys that force the decoder to reconstruct the original cover scene:
\begin{equation} 
\begin{split}
\mathcal{L}_\text{incorrect}=&(1-\lambda_\text{incorrect})\mathcal{L}_1(I_\text{pred\_incorrect},I_\text{gt\_cover})\\
&+\lambda\mathcal{L}_\text{SSIM}\left(I_\text{pred\_incorrect},I_\text{gt\_cover}\right).\label{eq:loss_incorrect}
\end{split}
\end{equation}
The newly introduced $\mathcal{L}_\text{incorrect}$ ensures that when the decoder is provided with an incorrect key, it reconstructs only the cover scene without revealing any hidden information. This enforces robustness by preventing unauthorized access and strengthens the security of the steganographic system. 

\vspace{0.3cm}
\noindent\textbf{The Feature-Contribution Exploration} aims to investigate the contribution of different 3D Gaussian attributes to secret hiding. The proposed decoder consists of a shared common branch and multiple feature-specific branches (\cref{fig_flowchart}(b)). The common branch captures comprehensive representations by leveraging the full 3D Gaussian feature space. Feature-specific decoder branches isolate distinct Gaussian attributes, enabling systematic quantification of each parameter's contribution to steganographic efficacy. Concretely, all 3D Gaussian features and the user-specific key are normalized and concatenated into: $\mathbf{f}=
\text{concat}\left(\alpha, \mathbf{r}, \mathbf{s}, \boldsymbol{\mu}, \mathbf{c},\mathbf{k}\right)$. Then $\mathbf{f}$ is input as the common branch $\text{MLP}_{\text{common}}$ to get the common feature $\mathbf{h}=\text{MLP}_{\text{common}}(\mathbf{f})$. The common feature $\mathbf{h}$ and the Gaussian features of cover $\{\alpha, \mathbf{r}, \mathbf{s}, \boldsymbol{\mu}, \mathbf{c}\}$ would be passed through feature-specific branches to obtain the updated Gaussian features $\{\alpha^*, \mathbf{r}^*, \mathbf{s}^*, \boldsymbol{\mu}^*, \mathbf{c}^*\}$:
\begin{equation}
     \begin{pmatrix} \alpha^*\\ \mathbf{r}^*\\ \mathbf{s}^*\\ \boldsymbol{\mu}^*\\ \mathbf{c}^* \end{pmatrix}=\begin{pmatrix} \alpha\\ \mathbf{r} \\ \mathbf{s} \\ \boldsymbol{\mu} \\ \mathbf{c} \end{pmatrix} +\boldsymbol{\theta}\circ\begin{pmatrix}
         \text{MLP}_{op}(\textbf{h})\\\text{MLP}_{ro}(\textbf{h})\\\text{MLP}_{sc}(\textbf{h})\\\text{MLP}_{po}(\textbf{h})\\\text{MLP}_{sh}(\textbf{h})\\
     \end{pmatrix},
\end{equation}
where $\text{MLP}_{op}$, $\text{MLP}_{ro}$, $\text{MLP}_{sc}$, $\text{MLP}_{po}$, and $\text{MLP}_{sh}$ represent feature-specific branches for opacity, rotation, scale, position, and SH features, respectively. 
$\boldsymbol{\theta}\in\mathbb{R}^5$ is a binary vector that enables the selection of update combinations for the corresponding features. $\circ$ is the Hadamard product. 

\subsection{3D-Sinkhorn Evaluation Metric}\label{sec_3dsinkhorn}
While fidelity can be quantified using standard image-space metrics such as PSNR, assessing steganographic security in 3D space requires a fundamentally different approach. We introduce a new security evaluation metric grounded in the Sinkhorn distance~\cite{sinkhorn_cuturi2013sinkhorn, sinkhorn2_feydy2019interpolating}, which measures the distributional disparities between original and steganographic 3D Gaussian parameters directly within the representation space.
This approach, inspired by recent advances in optimal transport for 3D applications~\cite{wast3d_kotovenko2024wast}, offers significant advantages over traditional 2D image-based metrics by detecting statistical anomalies in the underlying 3D representation that remain invisible in rendered views (\cref{fig_stegExpose}). The Sinkhorn distance provides an ideal balance between computational efficiency through entropic regularization and preservation of geometric correspondences critical for 3D analysis. During evaluation, 3D Gaussians from ground truth and stego cover scenes are normalized and projected into feature-specific histograms: ${g_i} ={\mathbf{hist}(f_i)}$, ${g_i^{gt}} ={\mathbf{hist}(f_i^{gt})}$, allowing us to quantify security across different attribute distributions:
\begin{equation}
    d = \sum_i(\mathbf{Sinkhorn}(g_i, g_i^{gt})).
\end{equation}  
By analyzing the distributional discrepancy between the ground truth cover and the stego cover in the 3D Gaussian parameter space, the 3D Sinkhorn distance metric quantifies steganographic imperceptibility at a fundamental level. Lower distributional discrepancy indicates that the stego cover maintains statistical properties nearly identical to the original, significantly enhancing resistance against both visual inspection and algorithmic detection methods. To systematically evaluate different feature combinations, we employ a composite score that balances reconstruction quality and statistical imperceptibility:
\begin{equation}
\text{score} = (\text{PSNR}_{\text{cover}} + \text{PSNR}_{\text{secret}}) \cdot (1-d).
\end{equation}
\subsection{Training Details}\label{sec_method_method}
\textbf{Dataset with Combined Camera Poses}: The training dataset consists of both ground truth cover images and hidden secret images, each paired with $\mathcal{M}$ corresponding camera poses. To augment our training dataset, we leverage the combined camera poses from both cover and secret scenes. For camera poses unique to either scene, we generate the corresponding ground truth images using pre-trained models: the cover 3D model (without embedding) for secret-scene poses, and vice versa. 

\vspace{0.3cm}
\noindent\textbf{Initialization with Combined SfM Points}: The initialization from the SfM point cloud is crucial for learning 3D Gaussians. Our method uniquely employs a combined SfM point cloud for initialization, preserving the spatial information of both cover and secret scenes. This strategy significantly enhances reconstruction quality by capturing the structural context of both datasets during initialization.

\vspace{0.3cm}
\noindent\textbf{Backpropagation with Tripled Losses}: 
Considering both fidelity and security, the overall loss of the proposed method can be summarized as:
\begin{equation}
\mathcal{L}=\beta_\text{cover}\mathcal{L}_\text{cover}+\sum_{s=1}^S\beta_\text{secret}^s\mathcal{L}_\text{secret}^s+\beta_\text{incorrect}\mathcal{L}_\text{incorrect},
\end{equation}
where $\beta_\text{cover}$, $\beta_\text{secret}^s$, and $\beta_\text{incorrect}$ balance the contribution of each loss term. 

\vspace{0.3cm}
\noindent\textbf{Refinement with Combined Densification}: During backpropagation, a combined densification strategy is employed by leveraging the view-space positional gradients from both the cover and secret scenes. This strategy guides the process of cloning or splitting large Gaussians in $\mathcal{G}_{\text{cover}}$, allowing for finer control over the representation and improving the accuracy of the hidden information while preserving the integrity of the cover scene.

\begin{table*}[t]
    \centering
    \resizebox{1.0\linewidth}{!}{
    \begin{tabular}{c|cccccccccc|c}
    \toprule
        \multirow{2}{*}{Methods} & \multirow{2}{*}{\shortstack{Scene\\Type}} & Bicycle & Bonsai & Room & Flowers & Treehill & Garden & Stump & Counter & Kitchen & \multirow{2}{*}{Average$\uparrow$} \\ 
        ~ & ~ & Bicycle & Bonsai & Room & Flowers & Treehill & Garden & Stump & Counter & Kitchen & ~ \\ \midrule
        3DGS-GT (GS-Hider) & cover  & 25.246 & 31.980 & 30.632 & 21.520 & 22.490 & 27.410 & 26.550 & 28.700 & 30.317 & 27.205 \\ \midrule
        3DGS-GT (KeySS) & cover  & 23.395 & 31.690 & 31.712 & 19.886 & 22.579 & 25.690 & 24.670 & 28.702 & 30.338 & 26.518 \\ \midrule\midrule
        \multirow{2}{*}{3DGS+SH} 
        & cover & 23.365 & 26.286 & 29.311 & 18.998 & 21.479 & 24.897 & 22.818 & 26.893 & 28.150 & 24.689 \\ 
        & \cellcolor[gray]{0.9}secret & \cellcolor[gray]{0.9}23.548 & \cellcolor[gray]{0.9}21.340 & \cellcolor[gray]{0.9}22.231 & \cellcolor[gray]{0.9}25.080 & \cellcolor[gray]{0.9}20.619 & \cellcolor[gray]{0.9}28.450 & \cellcolor[gray]{0.9}24.067 & \cellcolor[gray]{0.9}20.997 & \cellcolor[gray]{0.9}22.758 & \cellcolor[gray]{0.9}23.232 \\ \midrule
         \multirow{2}{*}{\shortstack{3DGS\\+Decoder}} 
         & cover & 23.914 & 27.674 & 27.502 & 19.877 & 21.200 & 24.284 & 24.134 & 26.561 & 26.013 & 24.573 \\
         & \cellcolor[gray]{0.9}secret & \cellcolor[gray]{0.9}20.611 & \cellcolor[gray]{0.9}20.318 & \cellcolor[gray]{0.9}21.668 & \cellcolor[gray]{0.9}20.540 & \cellcolor[gray]{0.9}19.848 & \cellcolor[gray]{0.9}25.287 & \cellcolor[gray]{0.9}19.933 & \cellcolor[gray]{0.9}20.670 & \cellcolor[gray]{0.9}22.367 & \cellcolor[gray]{0.9}21.249 \\ \midrule
        \multirow{2}{*}{\shortstack{GS-Hider\\\cite{gshider_zhang2024gshider}}} 
         & cover & 24.018 & 29.643 & 28.865 & 20.109 & 21.503 & 26.753 & 24.573 & 27.445 & 29.447 & 25.817 \\ 
         & \cellcolor[gray]{0.9}secret & \cellcolor[gray]{0.9}28.219 & \cellcolor[gray]{0.9}23.846 & \cellcolor[gray]{0.9}22.885 & \cellcolor[gray]{0.9}26.389 & \cellcolor[gray]{0.9}20.276 & \cellcolor[gray]{0.9}32.348 & \cellcolor[gray]{0.9}25.161 & \cellcolor[gray]{0.9}20.792 & \cellcolor[gray]{0.9}26.690 & \cellcolor[gray]{0.9}25.178 \\ \midrule
        \multirow{2}{*}{\shortstack{KeySS\\(op,ro,sc,xyz)}} 
         & cover & 23.011 & 31.081 & 30.785 & 19.476 & 22.433 & 25.225 & 23.827 & 28.120 & 29.862 & \textbf{25.980} \\
         & \cellcolor[gray]{0.9}secret & \cellcolor[gray]{0.9}29.533 & \cellcolor[gray]{0.9}25.456 & \cellcolor[gray]{0.9}23.877 & \cellcolor[gray]{0.9}29.272 & \cellcolor[gray]{0.9}22.121 & \cellcolor[gray]{0.9}28.179 & \cellcolor[gray]{0.9}29.452 & \cellcolor[gray]{0.9}20.891 & \cellcolor[gray]{0.9}29.064 & \cellcolor[gray]{0.9}\textbf{26.427} \\ \midrule 
        \multirow{2}{*}{\shortstack{KeySS\\(wrong key)}} 
         & v.s. cover & 22.959 & 31.023 & 30.719 & 19.625 & 22.443 & 25.120 & 23.763 & 28.098 & 29.495 & 25.916 \\ 
         & v.s. \cellcolor[gray]{0.9}secret & \cellcolor[gray]{0.9}11.010 & \cellcolor[gray]{0.9}10.190 & \cellcolor[gray]{0.9}9.751 & \cellcolor[gray]{0.9}9.624 & \cellcolor[gray]{0.9}12.458 & \cellcolor[gray]{0.9}9.843 & \cellcolor[gray]{0.9}11.228 & \cellcolor[gray]{0.9}10.373 & \cellcolor[gray]{0.9}8.593 & \cellcolor[gray]{0.9}10.341 \\ 
    \bottomrule
    \end{tabular}}
    \vspace{-0.1cm}
    \caption{PSNR scores for comparisons with previous works on single-secret hiding. 3DGS-GTs represent the ground truth 3DGS models used for training, serving as the theoretical upper bound for the performance of KeySS and GS-Hider, respectively. The results showcase the top 3 feature update combinations explored based on secret fidelity. 
    For wrong key inputs, PSNR scores are evaluated against cover (``vs. cover'') and secret (``vs. secret'') scenes to measure the effectiveness of unauthorized access prevention. Features are denoted as: opacity (op), rotation (ro), scale (sc), position (xyz), and SH (sh).}
 \label{tab_1hide1}
\end{table*}

\begin{figure*}[t]
\centering
\includegraphics[width=0.8\linewidth]{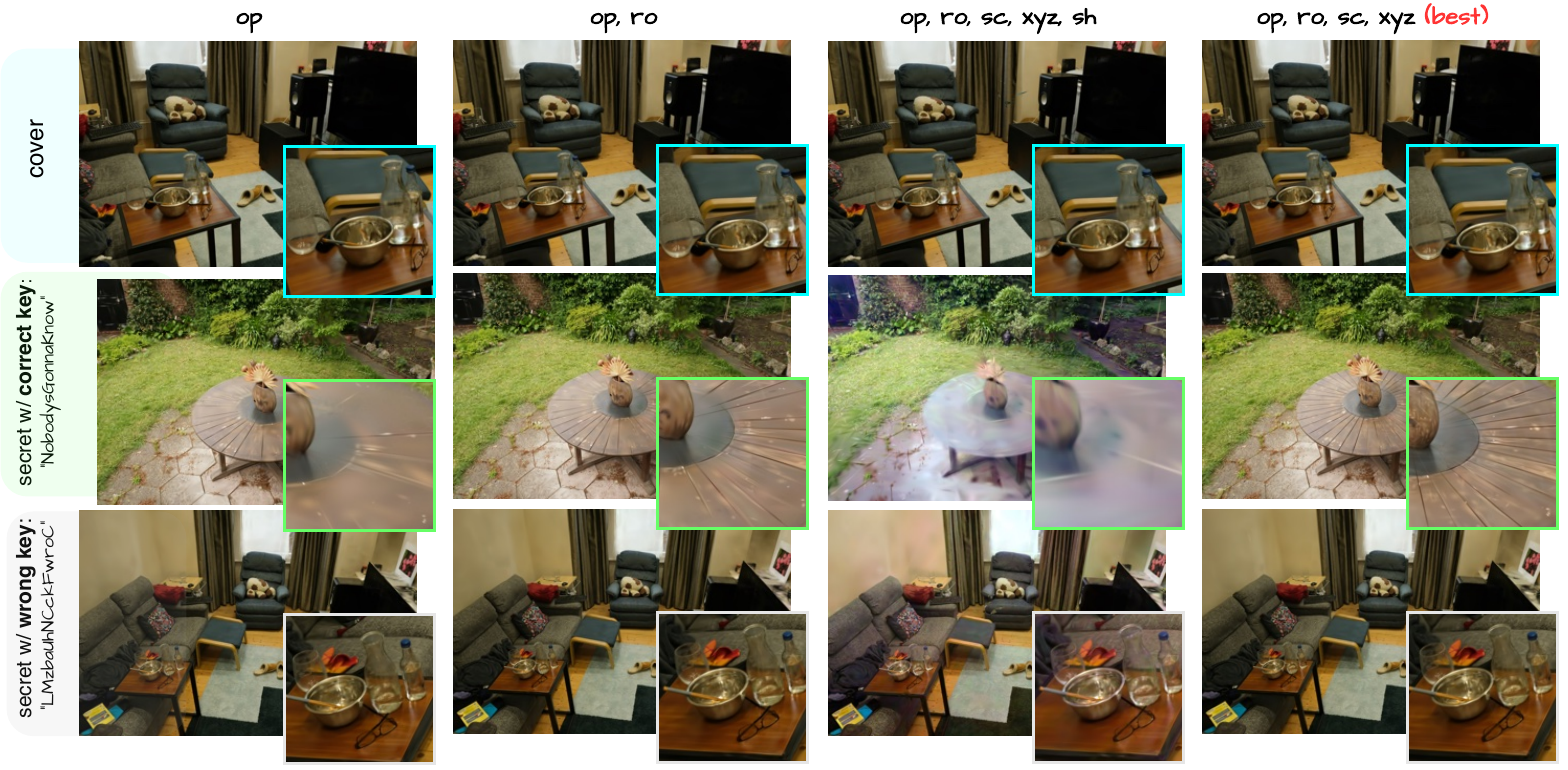}
\vspace{-0.15cm}
\caption{Visualization of decoder outputs across different feature combinations using correct and incorrect keys. The last two rows show secret recovery (correct key) and security preservation (incorrect key). Notation follows~\cref{tab_1hide1}.}
\vspace{-0.13cm}
\label{fig_1hid1vis}
\end{figure*}

\section{Experimental Results}
\begin{figure*}[ht]
\centering
\includegraphics[width=0.8\textwidth]{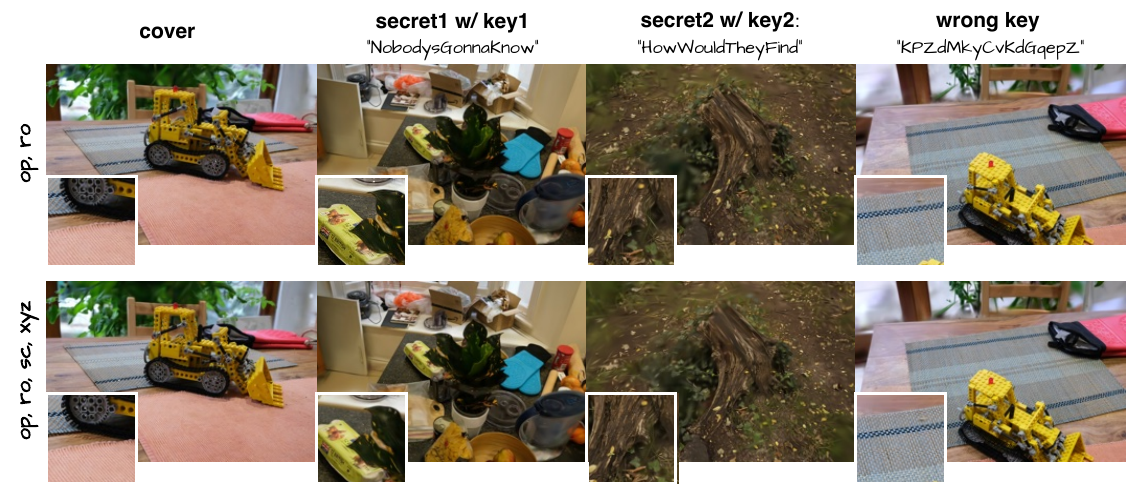}
\vspace{-0.2cm}
\caption{Visualization comparison of our method on multiple secret hiding across different feature update scenarios using both correct and incorrect key inputs. The notation is consistent with~\cref{tab_1hide1}. More visualization results can be found in~\cref{sec_sup_morevis}.}
\vspace{-0.2cm}
\label{fig_1hid2vis}
\end{figure*}
\subsection{Dataset and Implementation Details}
Following GS-Hider~\cite{gshider_zhang2024gshider}, 9 original scenes from the Mip-NeRF360 dataset~\cite{mipnerf360_barron2022mip} and 1 scene from the Deep Blending dataset~\cite{deep_blending_hedman2018deep} are paired into 9 cover-secret pairs (\cref{tab_1hide1}). To evaluate multi-secret hiding performance, we further create $(\text{cover},\text{secret}_1,\text{secret}_2)$ triplets from Mip-NeRF360 scenes (\cref{tab_1hide2}), showcasing our method's ability to embed and conceal multiple secrets within complex 3D environments.  Following the evaluation protocol of Mip-NeRF360, we use the test split of the dataset for assessment.

Our method is built upon the original 3DGS framework~\cite{3dgs_kerbl20233d} while maintaining full compatibility with recent advancements in 3D Gaussian Splatting. To maintain consistency, we adopt the same training hyper-parameters and rendering processes as the standard 3DGS~\cite{3dgs_kerbl20233d}, running for 30,000 iterations on a 24GB NVIDIA RTX 6000 GPU. Due to GPU memory constraints, we limit the number of Gaussians in the trained 3DGS cover scene to 500,000. Ground-truth scenes are also trained using the original 3DGS framework. The loss coefficients, $\lambda_\text{cover}$, $\lambda_\text{secret}$, and $\lambda_\text{incorrect}$, are empirically set to 0.5 following the original 3DGS settings. Additionally, the balance coefficients $\beta_\text{cover}$ and $\beta_\text{secret}$ are set to 0.5, while $\beta_\text{incorrect}$ is set to 0.01. These values are selected to maintain a balance between  cover reconstruction, secret recovery, and robustness against incorrect key attacks. For key embedding, we restrict the key space to 16-character alphanumeric strings (including both uppercase and lowercase letters), ensuring a balanced trade-off between security strength and computational feasibility. The user key is embedded with the pre-trained CLIP ViT-L/14 model. Performance evaluation employs the PSNR metric. SSIM and LPIPS analyses and more detailed analysis are available in Appendix.

\subsection{Fidelity Assessment}
\textbf{Single-Secret Hiding}: We compare our KeySS method against the baseline GS-Hider~\cite{gshider_zhang2024gshider}, as shown in~\cref{tab_1hide1}.
As the open-source implementations of the existing methods are unavailable, we directly compare our results with the reported figures from their papers. \cref{tab_1hide1} first presents the quality of our ground truth 3DGS cover, which is trained from scratch~\cite{3dgs_kerbl20233d} and serves as the theoretical upper bound of our performance. However, when compared to the ground truth 3DGS used in GS-Hider, our upper bound is lower, indicating differences in the underlying training setups.
Despite having a lower upper limit, our KeySS method achieves higher fidelity in cover reconstruction, with a minimal fidelity reduction of only 0.511 dB from the upper limit, compared to 1.388 dB in the baseline method.
KeySS also excels in secret preservation, outperforming the baseline by $4.9\%$. The proposed end-to-end framework is simple yet powerful in achieving high-quality secret hiding. \vspace{0.3cm}
\begin{figure}[t]
\centering
\vspace{-0.1cm}
\includegraphics[width=0.95\linewidth]{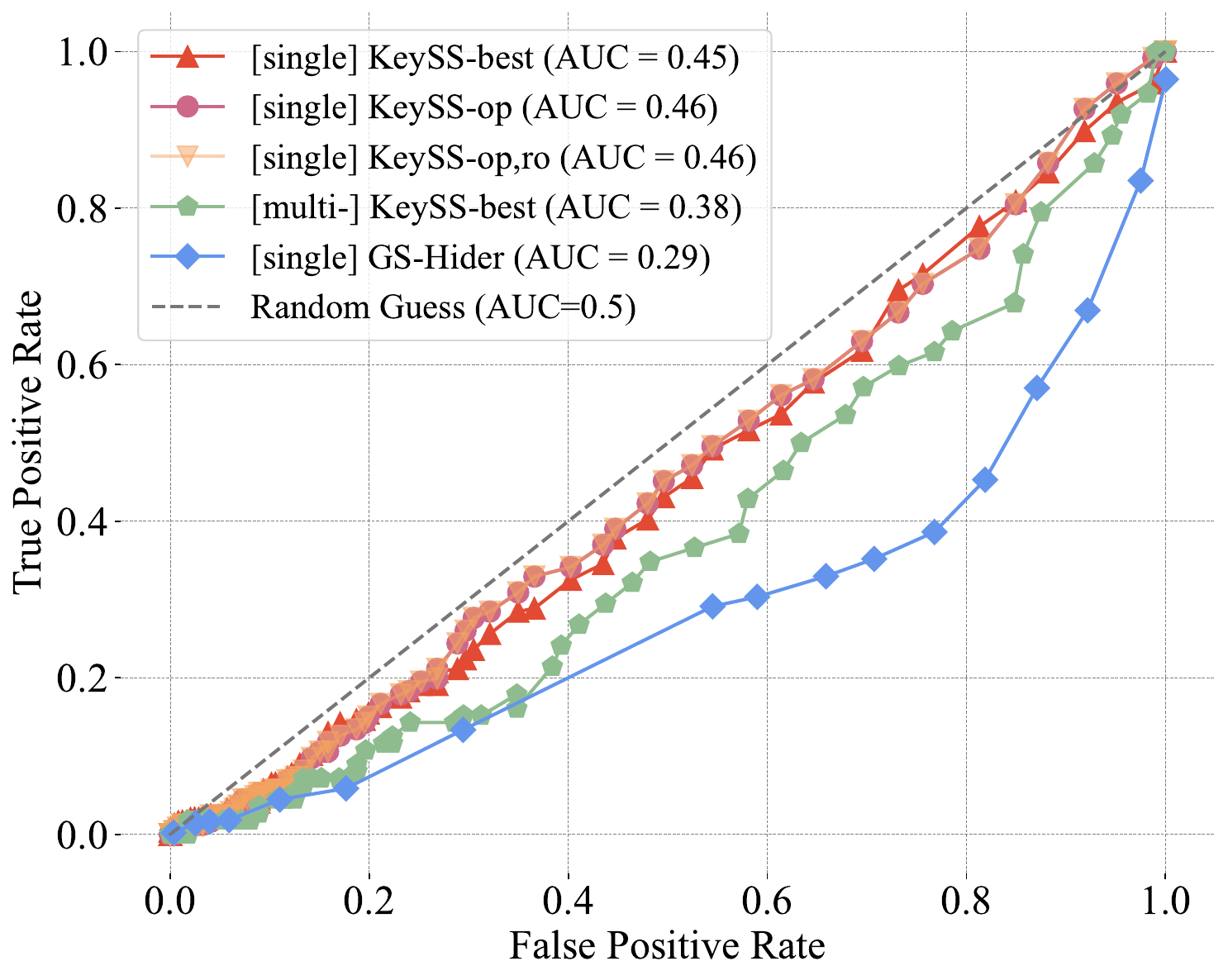}
\vspace{-0.4cm}
\caption{ROC curves from StegExpose analysis. `single' and `multi-' refer to single- and multi- secret hiding, respectively. `best' denotes the optimal feature combination (op, ro, sc, xyz).}
\label{fig_stegExpose}
\vspace{-0.2cm}
\end{figure}

\noindent\textbf{Multiple-Secret Hiding}: To seamlessly embed an additional secret into the cover scene, we extend our model by simply incorporating another secret loss term, without introducing any further architectural modifications. This straigtforward extension highligts the flexibility of our approach, as it naturally scales to multiple hidden secrets without requiring structural changes or additional constraints.~\cref{tab_1hide2} presents the PSNR scores for multi-secret hiding, demonstrating that our method preserves high cover scene fidelity while maintaining competitive secret reconstruction quality, comparable to the single-secret results in~\cref{tab_1hide1}. As illustrated in~\cref{fig_1hid2vis}, our model effectively reveals different hidden secrets based on the provided key input. This demonstrates the robustness and adaptability of our key-controllable decoding mechanism, ensuring that each unique key accurately retrieves its corresponding secret while maintaining the integrity of the cover scene. 

\subsection{Security Evaluation} \label{sec_exp_security}
\textbf{2D Steganalysis with StegExpose}: Data security is a fundamental concern in steganography, where the goal is to conceal information without raising suspicion. The resilience against steganalysis attacks is evaluated using advanced 2D steganalysis tools, specifically StegExpose~\cite{stegexpose_boehm2014stegexpose}, to analyze the detectability of stego cover scenes.
A detection dataset is constructed by mixing an equal proportion of cover scenes with and without hidden secrets. The receiver operating characteristic (ROC) curves in~\cref{fig_stegExpose} are generated by varying StegExpose detection thresholds across a broad range. In an ideal scenario, the steganalyzer should perform no better than random guessing ($50\%$ classification probability, $\text{AUC}=0.5$), resulting in an ROC curve along the diagonal~\cite{hinet_jing2021hinet,video_in_video_mou2023large}.
As shown in~\cref{fig_stegExpose}, the proposed method demonstrates superior resistance to steganalysis attacks, producing significantly less detectable stego covers compared to existing approaches. Notably, this strong security property is maintained even when embedding multiple secrets, highligting the framework's effectiveness in achieving secure and imperceptible information hiding. 
Moreover, traditional 2D image-based metrics are inadequate for evaluating the security of 3D steganography. As shown in~\cref{fig_stegExpose}, different feature update strategies in our method yield similar StegaExpose results (similar AUC values), failing to accurately reflect the robustness of the 3DGS models. This highlights the need for dedicated 3D-specific evaluation metrics that can effectively capture the spatial and structural imperceptibility of hidden information within 3D Gaussian splatting (\cref{sec_3dsinkhorn}).

\vspace{0.3cm}
\begin{table}[t]
    \centering
    \resizebox{\linewidth}{!}{
    \begin{tabular}{l|lc|lc|lc}
    \toprule
        Type & Scenes & PSNR$\uparrow$ & Scenes & PSNR$\uparrow$ & Scenes & PSNR$\uparrow$ \\ \midrule
        Cover & Flower & 18.990 & Bicycle & 22.478 & Kitchen & 30.103 \\ 
        \cellcolor[gray]{0.9}Secret1 & \cellcolor[gray]{0.9}Treehills & \cellcolor[gray]{0.9}21.038 & \cellcolor[gray]{0.9}Bonsai & \cellcolor[gray]{0.9}29.031 & \cellcolor[gray]{0.9}Counter & \cellcolor[gray]{0.9}26.753 \\ 
        \cellcolor[gray]{0.8}Secret2 & \cellcolor[gray]{0.8}Garden & \cellcolor[gray]{0.8}21.870 & \cellcolor[gray]{0.8}Room & \cellcolor[gray]{0.8}26.544 & \cellcolor[gray]{0.8}Stump & \cellcolor[gray]{0.8}21.519 \\ \midrule\midrule
        \multirow{2}{*}{\shortstack{Wrong\\Key}} & vs.Cover & {18.778} & vs.Cover & {16.456} & vs.Cover & 29.638 \\ 
        ~ & vs.Secret1 & 10.598 & vs.Secret1 & 12.117 & vs.Secret1 & 8.910 \\ \bottomrule
    \end{tabular}}
    \vspace{-0.2cm}
 \caption{PSNR scores for multiple-secret hiding performance.} 
 \label{tab_1hide2}
 \vspace{-0.2cm}
\end{table}

\noindent\textbf{Security Against Unauthorized Access:} Our proposed decoder employs a key-controllable scheme, effectively defending against wrong key attacks. The last row of~\cref{tab_1hide1} presents the performance of the $\mathcal{L}_{\text{incorrect}}$ loss in preventing unauthorized secret retrieval. For this evaluation, we randomly generate incorrect keys that were never encountered during training. The average PSNR scores against both the ground truth cover and secret scenes demonstrate the robustness of our decoder in safeguarding the hidden information.
Furthermore,~\cref{tab_1hide2} extends this evaluation to multiple-secret hiding scenarios, reinforcing the effectiveness of our method in maintaining security across different settings. The visual examples in~\cref{fig_1hid1vis} and~\cref{fig_1hid2vis} illustrate how the decoder reacts to incorrect key inputs, ensuring that secret content remains inaccessible without the correct key. 
The key-controllable scheme demonstrates both flexibility and efficiency, readily incorporating additional loss for enhanced security against unauthorized access.

\subsection{Ablation Study}
\begin{figure}[t]
\centering
\includegraphics[width=\linewidth]{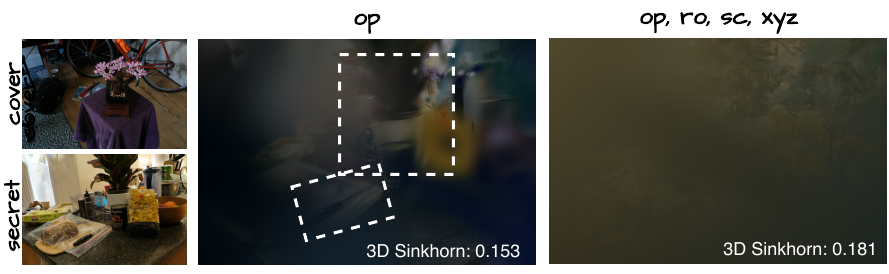}
\vspace{-0.6cm}
\caption{Visualize low-opacity areas of the cover image}
\vspace{-0.2cm}
\label{fig_lowopacity}
\end{figure}

\textbf{Balancing Fidelity and Security}: 
The ablation study in~\cref{tab_ablation_sinkhorn} demonstrates the performance of different feature update combinations learned by our decoder. The top three combinations, based on cover and secret fidelity, are: a single feature update with opacity, a double feature update with opacity and rotation, and a quadruple-feature update with opacity, rotation, scale, and position. 
Detailed PSNR results for the top three feature combinations are presented in~\cref{tab_1hide1}.
Interestingly, the results align to our intuition (\cref{fig_motivation}), the opacity feature appears to contribute the most to secret hiding. In contrast, the color/SH feature, due to its complexity and high-dimensional nature, is more difficult to learn and assist the concealment process in our method. 
The composite score reveals that updating four features (opacity, rotation, scale, and position) yields the best trade-off between reconstruction fidelity and security.
To further validate security,~\cref{fig_lowopacity} analyzes the distribution of low opacity areas in cover scenes for potential information leakage patterns. Specifically, we visualize suspicious regions by setting the opacity of cover 3D Gaussians below 0.05 to 1 and all others to zero. The quadruple-feature update demonstrates enhanced steganographic imperceptibility with the lower 3D-Sinkhorn distance. In contrast, updates using opacity alone exhibits compromised security due to more concentrated and potentially detectable patterns in the low-opacity regions. 
\begin{table}[!t]
\vspace{-0.2cm}
    \centering
    \resizebox{\linewidth}{!}{
    \begin{tabular}{c|ccccc|cc|c|c}
    \toprule
        \multirow{2}{*}{Count} & \multicolumn{5}{c|}{Updated Features}  & \multicolumn{2}{c|}{Scene PSNR} & \multicolumn{1}{c|}{\multirow{2}{*}{\shortstack{3D\\Sinkhorn$\downarrow$}}} & \multirow{2}{*}{\shortstack{Score$\uparrow$}} \\ \cmidrule{2-8}
        ~ & \multicolumn{1}{c|}{op} & \multicolumn{1}{c|}{ro} & \multicolumn{1}{c|}{sc} & \multicolumn{1}{c|}{xyz} & \multicolumn{1}{c|}{sh} & \multicolumn{1}{c|}{Cover$\uparrow$} & \multicolumn{1}{c|}{Secret$\uparrow$} & \multicolumn{1}{c|}{~} &~ \\ \midrule
        1 & \cellcolor[gray]{0.9}\faCheck & \cellcolor[gray]{0.9}~ & \cellcolor[gray]{0.9}~ & \cellcolor[gray]{0.9}~ & \cellcolor[gray]{0.9}~ & \cellcolor[gray]{0.9}26.020 & \cellcolor[gray]{0.9}26.138 & \cellcolor[gray]{0.9}0.181 & \cellcolor[gray]{0.9}42.717 \\
        ~ & ~ & \faCheck & ~ & ~ & ~ & 21.752 & 23.784 & 0.268 & 33.314 \\ 
        ~ & ~ & ~ & \faCheck & ~ & ~ & 25.080 & 20.497 & 0.190 & 36.921 \\ 
        ~ & ~ & ~ & ~ & \faCheck & ~ & 21.980 & 23.866 & 0.211 & 36.182 \\ 
        ~ & ~ & ~ & ~ & ~ & \faCheck & 22.825 & 12.530 & 0.162 & 29.621 \\\midrule
        2 & \cellcolor[gray]{0.8}\faCheck & \cellcolor[gray]{0.8}\faCheck & \cellcolor[gray]{0.8}~ & \cellcolor[gray]{0.8}~ & \cellcolor[gray]{0.8}~ & \cellcolor[gray]{0.8}26.036 & \cellcolor[gray]{0.8}26.113 & \cellcolor[gray]{0.8}0.175 & \cellcolor[gray]{0.8}43.008 \\ 
        ~ & \faCheck & ~ & \faCheck & ~ & ~ & 25.998 & 26.038 & 0.208 & 41.228 \\ 
        ~ & \faCheck & ~ & ~ & \faCheck & ~ & 25.777 & 21.615 & 0.329 & 31.795 \\ 
        ~ & \faCheck & ~ & ~ & ~ & \faCheck & 25.831 & 24.118 & 0.213 & 39.319 \\ \midrule
        3 & \faCheck & \faCheck & \faCheck & ~ & ~ & 25.632 & 24.743 & 0.206 & 39.987 \\ 
        ~ & \faCheck & \faCheck & ~ & \faCheck & ~ & 25.411 & 25.662 & 0.219 & 39.899 \\  
        ~ & \faCheck & \faCheck & ~ & ~ & \faCheck & 24.643 & 21.645 & 0.185 & 37.729 \\ \midrule
        4 & \cellcolor[gray]{0.7}\faCheck & \cellcolor[gray]{0.7}\faCheck & \cellcolor[gray]{0.7}\faCheck & \cellcolor[gray]{0.7}\faCheck & \cellcolor[gray]{0.7}~ & \cellcolor[gray]{0.7}25.980 & \cellcolor[gray]{0.7}26.427 & \cellcolor[gray]{0.7}0.153 & \cellcolor[gray]{0.7}\textbf{44.389} \\ 
        ~ & \faCheck & \faCheck & ~ & \faCheck & \faCheck & 25.951 & 19.970 & 0.430 & 26.157 \\  \midrule
        5 & \faCheck & \faCheck & \faCheck & \faCheck & \faCheck & 25.832 & 20.961 & 0.256 & 34.810 \\ \bottomrule
    \end{tabular}}
    \vspace{-0.1cm}
    \caption{Ablation study on different feature update combinations. Top three methods highlighted in gray.}
    \vspace{-0.15cm}
    \label{tab_ablation_sinkhorn}
\end{table}
\begin{table}[!t]
\vspace{-0.2cm}
    \centering
    \resizebox{1\linewidth}{!}{
    \begin{tabular}{c|c|c|ccc}
    \toprule
        \shortstack{Combined} & \shortstack{Combined} & \shortstack{Combined} & \multirow{2}{*}{Cover} & \multirow{2}{*}{Secret} & \multirow{2}{*}{Average} \\ 
        \shortstack{SfM} & \shortstack{Camera Poses} & \shortstack{Densification} & ~ & ~ & ~ \\ \midrule
        cover & cover & \faCheck & 22.830 & 29.719 & 26.275 \\ 
        cover & \faCheck & \faCheck & 23.279 & 27.083 & 25.181 \\ 
        \faCheck & cover & \faCheck & 23.300 & 26.855 & 25.077 \\ 
        secret & \faCheck & \faCheck & 19.907 & 29.761 & 24.834 \\ 
        secret & secret & \faCheck & 19.789 & 29.719 & 24.754 \\ 
        \faCheck & secret & \faCheck & 19.763 & 29.693 & 24.728 \\ 
        \faCheck & \faCheck & cover & 22.888 & 29.611 & 26.249 \\ 
        \faCheck & \faCheck & \faCheck & 22.911 & 29.858 & \textbf{26.384} \\ \bottomrule
    \end{tabular}}
    \vspace{-0.15cm}
    \caption{Ablation study on combined SfM initialization, camera poses, and densification.}
    \label{tab_exp_combine}
    \vspace{-0.25cm}
\end{table}

\noindent\textbf{Combined Gaussian Optimization}: The proposed method integrates three combination strategies: combined SfM point clouds for initialization, combined camera poses for training sample enrichment, and combined densification for refinement. The ablation study results are summarized in~\cref{tab_exp_combine}. The SfM initialization proves crucial for reconstruction performance, as using only cover or secret scenes' SfM points as initial 3D Gaussians results in biased performance toward the respective scene. The combined camera poses strategy achieves more balanced performance across both cover and secret scenes by providing diverse training samples. Furthermore, combined densification enhances performance by leveraging view-space positional gradients from both scenes during refinement.

\section{Conclusion and Limitations}
In this paper, we introduce an end-to-end 3D steganography framework, KeySS, that simultaneously optimizes the cover 3D Gaussians and a key-secured decoder. 
Our decoder preserves imperceptibility by adhering to the standard 3D-GS format and rendering pipeline while incorporating a key-controllable scheme, enabling robust multi-secret hiding and resilience against incorrect key attacks. 
Furthermore, task-specific branches in the decoder enable the systematic exploration of the optimal feature update for high-fidelity secret concealment. We also introduce 3D-Sinkhorn, designed to quantify steganographic imperceptibility, overcoming the limitations of traditional 2D steganalysis metrics and laying a foundation for future research in 3D steganography.
Extensive experimental results demonstrate that KeySS achieves state-of-the-art performance in both fidelity and security, validating its effectiveness for secure 3D information embedding. 

\vspace{0.4cm}
\noindent\textbf{Limitations}: While achieving high performance in 3D steganography, the method faces an inherent trade-off between cover and secret fidelity due to joint optimization of 3DGS and key-secured decoder (\cref{tab_ablation_sinkhorn}). This stems from the challenge of simultaneously optimizing cover rendering and secret embedding within shared Gaussian features. Further exploration of loss balancing and feature modulation could potentially mitigate this issue.
\clearpage
{
    \small
    \bibliographystyle{ieeenat_fullname}
    \bibliography{main}
}

\clearpage
\setcounter{page}{1}
\maketitlesupplementary
\appendix
\renewcommand{\thesubsection}{\Alph{subsection}}
Additional implementation details and comparisons are provided in the supplementary material. Unless stated otherwise, all experiments are conducted using our method with the optimal feature update combination: (op,ro,sc,xyz).
\subsection{Related Work on 3D Scene Reconstruction}\label{sec_sup_related_work}
3D scene reconstruction aims to generate a 3D scene from a set of images and other data, while rendering projects 3D models into 2D images based on given camera poses. Traditional methods include structure-from-motion (SfM)~\cite{sfm_snavely2006photo} and multi-view stereo (MVS)~\cite{sfm_snavely2006photo} algorithms. With the rise of deep learning, NeRF~\cite{nerf_mildenhall2021nerf} encodes scene information by overfitting a multi-layer perceptron (MLP), enabling photorealistic novel view synthesis from limited input images. Despite revolutionizing image synthesis, NeRF suffers from high computational costs and limited controllability due to its implicit representation.
3DGS~\cite{3dgs_kerbl20233d} emerges as a solution to these challenges, providing an explicit representation and highly parallelized workflows for efficient rendering and reconstruction. It represents scenes with learnable 3D Gaussians, which are projected onto image planes through a splatting process, enabling high-quality rendering with real-time performance. Our method builds upon the strengths of 3DGS in both reconstruction and rendering while leveraging the high-capacity embedding potential of millions of 3D Gaussians to conceal the secret scenes effectively.
\subsection{Implementation Details}
\label{app:imp}
The proposed decoder (\cref{fig_flowchart} (b)) consists of a shared common branch and multiple feature-specific branches, ensuring both simplicity and efficiency. The common branch comprises two MLP layers with ReLU activation for general feature extraction. The common branch receives concatenated features derived from L2-normalized Gaussian attributes as input.
Each feature-specific branch contains two additional MLP layers followed by feature-specific activation functions. To enhance training stability and facilitate gradient flow, residual connections are incorporated. 
For the MLP architecture, we employ 1×1 convolutional layers instead of standard linear layers, which is motivated by their effectiveness in PointNet~\cite{PointNet_qi2017pointnet}. 
The use of 1×1 convolutions enables efficient local feature aggregation while maintaining spatial awareness, leading to improved performance in our feature update strategy.
Based on this architecture, the proposed decoder is equipped with a key-controllable mechanism for enhanced security and a selective update scheme to fully explore the hiding potential of various feature update combinations.\newline

\subsection{Rendering Speed Comparison}
As shown in~\cref{tab_sup_fps}, we assess the adaptability of our method and baselines within the SIBR Viewer rendering engine provided by 3DGS~\cite{3dgs_kerbl20233d}. Unlike GSHider, but similar to WaterGS, our approach maintains full compatibility with the standard 3DGS format and rendering process, allowing seamless integration into the original 3DGS pipeline without requiring modifications. This ensures that our method can be readily deployed in existing 3DGS-based applications without additional engineering overhead.

Furthermore,~\cref{tab_sup_fps} demonstrates that our method preserves the standard rendering efficiency of 3DGS, achieving an average rendering speed of 130 FPS. This result indicates that our steganographic enhancements do not introduce significant computational overhead, maintaining real-time performance comparable to the original 3DGS framework.

\subsection{Detailed 3D-Sinkhorn Results} \label{sup_table1}
A comprehensive breakdown of the 3D Sinkhorn distance analysis, including per-attribute comparisons, is presented in~\cref{tab_sup_3dsinkhorn}. This metric provides a fine-grained evaluation of the distributional discrepancy between the ground truth cover and the stego cover in the 3D Gaussian space, offering crucial insights into the imperceptibility of the hidden secret. As we can find, the histogram distance between the scales' histograms are very close which indicates that the scale value tends to distribe relatively equally in 3DGS models. The other feature distribution differs more.

A comprehensive breakdown of the 3D-Sinkhorn distance analysis, including detailed per-attribute comparisons across opaciy, rotation, scale, position, and spherical harmonics, is presented in~\cref{tab_sup_3dsinkhorn}. This sophisticated metric quantifies the distributional discrepancy between the ground truth cover and the stego cover in the 3D Gaussian parameter space, offering crucial insights into the imperceptibility of the hidden secret. The analysis reveals a notable pattern: the histogram distances between scale distributions are consistently minimal across all tested scenes, indicating that scale parameters tend to distribute relatively uniformly in well-optimized 3DGS models regardless of content. In contrast, other feature distributions, particularly opacity and rotation, exhibit higher variability between cover and stego models. This disparity suggests that these attributes provide more exploitable degrees of freedom for secret embedding while maintaining perceptual fidelity, aligning with our quantitative performance results. The 3D-Sinkhorn analysis thus provides statistical validation for our feature combination strategy, confirming that the method optimally utilizes the available feature space for steganographic purposes.

\subsection{Additional Quantitative Results}\label{sec_sup_ssim}
Due to the unavailability of public codebases for existing 3D steganography methods~\cite{gshider_zhang2024gshider,sis_guo2024splats}, we independently train the ground-truth cover and secret scenes from scratch using the original 3DGS framework~\cite{3dgs_kerbl20233d}. This approach ensures a fair baseline for comparison while maintaining consistency with standard 3DGS optimization procedures and rendering pipelines. The quality metrics (PSNR, SSIM, and LPIPS) of the original scenes are comprehensively documented in~\cref{tab_sup_3dgs_gt}. Compared with the original scene performance reported by GS-Hider, our baseline models show lower scores across all 3 metrics. This performance gap in the ground truth models should be considered when interpreting the steganographic results, as it indicates our method starts from a lower baseline across these metrics.

Comprehensive quality metrics (PSNR, SSIM, and LPIPS) for our method with the optimal feature combination are presented in~\cref{tab_sup_ssim}. The results demonstrate that our approach achieves strong rendering performance, with only a minimal PSNR reduction of 0.78 dB compared to the original 3DGS baseline in~\cref{tab_sup_3dgs_gt}. While our SSIM and LPIPS scores appear lower than those reported by GS-Hider, this discrepancy primarily stems from our ground truth models starting from a lower baseline across these metrics. When evaluating the relative performance degradation from their respective baselines, our method demonstrates comparable reduction margins to GS-Hider. This suggests that despite operating under more challenging baseline conditions and incorporating additional security features, our steganographic approach preserves visual quality at a similar level. As previously mentioned, our method can be extended to other advanced 3DGS models, ensuring broader applicability and compatibility with recent developments in the field.

\subsection{Robustness of the Decoder}
\begin{table}[!t]
    \centering
    \resizebox{\linewidth}{!}{
    \begin{tabular}{c|c|c|c}
    \toprule
        Methods  & Scene Type & SIBR Viewer & Averaged Rendering FPS \\ \midrule
        3DGS~\cite{3dgs_kerbl20233d}&cover&\faCheck&130\\\midrule
        \multirow{2}{*}{GS-Hider~\cite{gshider_zhang2024gshider}} & cover & \faTimes & 45 \\ 
        ~ & secret & \faTimes & 45 \\ \midrule
        \multirow{2}{*}{WaterGS~\cite{sis_guo2024splats}} & cover & \faCheck & 100+ \\ 
        ~ & secret & \faCheck & 100+ \\ \midrule
        \multirow{2}{*}{KeySS} & cover & \faCheck & 130 \\
        ~ & secret & \faCheck & 130 \\ \bottomrule
    \end{tabular}}
    \caption{Comparison of the average rendering speed and the adaptability of SIBR viewer~\cite{3dgs_kerbl20233d}.}
    \label{tab_sup_fps}
\end{table}

\begin{table}[!t]
	\centering
	\resizebox{\linewidth}{!}{
		\begin{tabular}{c|ccccc|c|ccccc}
			\hline
			\multirow{2}{*}{Count} & \multicolumn{5}{c|}{Updated features} & \multicolumn{6}{c}{3D Sinkhorn Distance$\downarrow$} \\ \cmidrule{2-12}
			~ & op & ro & sc & xyz & SH & $\sum$ & op & sc & ro & xyz & SH \\ \midrule
			1 & \faCheck & ~ & ~ & ~ & ~ & 0.181 & 0.079 & 4e-4 & 0.030 & 0.038 & 0.034 \\ 
			~ & ~ & \faCheck & ~ & ~ & ~ & 0.268 & 0.133 & 5e-4 & 0.050 & 0.042 & 0.043 \\ 
			~ & ~ & ~ & \faCheck & ~ & ~ & 0.190 & 0.060 & 5e-4 & 0.041 & 0.042 & 0.047 \\
			~ & ~ & ~ & ~ & \faCheck & ~ & 0.211 & 0.099 & 4e-4 & 0.031 & 0.046 & 0.034 \\ 
			~ & ~ & ~ & ~ & ~ & \faCheck & 0.162 & 0.066 & 4e-4 & 0.030 & 0.040 & 0.026 \\ \midrule
			2 & \faCheck & \faCheck & ~ & ~ & ~ & 0.175 & 0.063 & 5e-4 & 0.038 & 0.034 & 0.040 \\ 
			~ & \faCheck & ~ & \faCheck & ~ & ~ & 0.208 & 0.093 & 5e-4 & 0.038 & 0.044 & 0.033 \\ 
			~ & \faCheck & ~ & ~ & \faCheck & ~ & 0.329 & 0.084 & 4e-4 & 0.101 & 0.073 & 0.071 \\ 
			~ & \faCheck & ~ & ~ & ~ & \faCheck & 0.213 & 0.095 & 4e-4 & 0.042 & 0.041 & 0.034 \\ \midrule
			3 & \faCheck & \faCheck & \faCheck & ~ & ~ & 0.206 & 0.084 & 5e-4 & 0.033 & 0.054 & 0.036 \\ 
			~ & \faCheck & \faCheck & ~ & \faCheck & ~ & 0.219 & 0.102 & 5e-4 & 0.048 & 0.043 & 0.026 \\
			~ & \faCheck & \faCheck & ~ & ~ & \faCheck & 0.185 & 0.077 & 4e-4 & 0.030 & 0.041 & 0.037 \\ \midrule
			4 & \faCheck & \faCheck & \faCheck & \faCheck & ~ & 0.153 & 0.052 & 5e-4 & 0.027 & 0.037 & 0.036 \\ 
			~ & \faCheck & \faCheck & ~ & \faCheck & \faCheck & 0.430 & 0.301 & 5e-4 & 0.038 & 0.038 & 0.053 \\ \midrule 
			5 & \faCheck & \faCheck & \faCheck & \faCheck & \faCheck & 0.256 & 0.125 & 5e-4 & 0.035 & 0.037 & 0.058 \\ \bottomrule
	\end{tabular}}
	\caption{Detailed breakdown of 3D-Sinkhorn distances.}
	\label{tab_sup_3dsinkhorn}
\end{table}

\begin{table}[!t]
\vspace{-0.2cm}
    \centering
    \resizebox{\linewidth}{!}{
    \begin{tabular}{c|c|c|c|c|c|c}
    \toprule
        \multirow{2}{*}{Scenes} & \multicolumn{3}{c|}{GS-Hider} & \multicolumn{3}{c}{KeySS (op,ro,sc,xyz)}\\ \cmidrule{2-7}
        ~ & PSNR$\uparrow$ & SSIM$\uparrow$ & LPIPS$\downarrow$ & PSNR$\uparrow$ & SSIM$\uparrow$ & LPIPS$\downarrow$\\ \midrule
        Bicycle & 24.377 & 0.735 & 0.254 & 23.333 & 0.596 & 0.424 \\ 
        Bonsai & 31.147 & 0.937 & 0.191 & 31.294 & 0.934 & 0.164 \\ 
        Room & 30.190 & 0.918 & 0.209 & 31.360 & 0.912 & 0.177 \\ 
        Flowers & 20.897 & 0.583 & 0.347 & 19.872 & 0.460 & 0.498 \\ 
        Treehill & 21.952 & 0.625 & 0.361 & 22.510 & 0.580 & 0.449 \\ 
        Garden & 26.954 & 0.855 & 0.118 & 25.683 & 0.757 & 0.284 \\ 
        Stump & 25.565 & 0.731 & 0.259 & 24.548 & 0.656 & 0.409 \\ 
        Counter & 28.053 & 0.899 & 0.202 & 28.433 & 0.887 & 0.189 \\ 
        Kitchen & 29.588 & 0.921 & 0.129 & 30.226 & 0.918 & 0.146 \\ \midrule
        Average & 26.525 & 0.800 & 0.230 & 26.362 & 0.744 & 0.304 \\ \hline
    \end{tabular}}
    \caption{PSNR, SSIM, and LPIPS scores of the pretrained ground-truth 3DGS models used in the proposed method, compared with those used in GS-Hider.}
    \label{tab_sup_3dgs_gt}
\end{table}
\begin{table}[!t]
    \centering
    \resizebox{\linewidth}{!}{
    \begin{tabular}{c|ccc|ccc}
    \toprule
        Cover& $\text{PSNR}_c\uparrow$ & $\text{SSIM}_c\uparrow$ & $\text{LPIPS}_c\downarrow$ & $\text{PSNR}_s\uparrow$ & $\text{SSIM}_s\uparrow$ & $\text{LPIPS}_s\downarrow$ \\ \midrule
        Bicycle & 23.011 & 0.566 & 0.457 & 29.533 & 0.895 & 0.293 \\ 
        Bonsai & 31.081 & 0.934 & 0.167 & 25.456 & 0.798 & 0.330 \\ 
        Room & 30.785 & 0.906 & 0.192 & 23.877 & 0.657 & 0.389 \\ 
        Flowers & 19.476 & 0.426 & 0.535 & 29.272 & 0.891 & 0.303 \\ 
        Treehill & 22.433 & 0.567 & 0.473 & 22.122 & 0.528 & 0.495 \\ 
        Garden & 25.225 & 0.725 & 0.324 & 28.179 & 0.884 & 0.320 \\
        Stump & 23.827 & 0.615 & 0.461 & 29.452 & 0.894 & 0.298 \\ 
        Counter & 28.120 & 0.883 & 0.200 & 20.892 & 0.467 & 0.548 \\ 
        Kitchen & 29.862 & 0.913 & 0.157 & 29.064 & 0.912 & 0.207 \\ \midrule
        \shortstack{Average\\KeySS} & 25.980 & 0.726 & 0.330 & 26.427 & 0.770 & 0.354 \\ \midrule
        \shortstack{Average\\GS-Hider} & 25.817 & 0.782 & 0.246 & 25.178 & 0.780 & 0.306 \\ \bottomrule
    \end{tabular}}
    \caption{All Metrics of our method on single secret hiding. Due to space constraints, secret scene names are omitted but can be found in~\cref{tab_1hide1}. Note that $\text{PSNR}_c$, $\text{SSIM}_c$, and $\text{LPIPS}_c$ respectively are used to evaluate the fidelity of the cover scenes, while $\text{PSNR}_s$, $\text{SSIM}_s$, and $\text{LPIPS}_s$ are for the fidelity of the secret scenes.}
    \label{tab_sup_ssim}
\end{table}
\begin{table}[t]
\vspace{-0.2cm}
    \centering
    \resizebox{\linewidth}{!}{
    \begin{tabular}{c|c|c|c|c}
    \toprule
        \multirow{2}{*}{Scene}~&\multicolumn{2}{c|}{KeySS w$/$o key} & \multicolumn{2}{c}{KeySS w$/$ key} \\ \cmidrule{2-5}
        ~ & $\text{PSNR}_{cover}\uparrow$ & $\text{PSNR}_{secret}\uparrow$ & $\text{PSNR}_{cover}\uparrow$ & $\text{PSNR}_{secret}\uparrow$ \\ \midrule
        Bicycle & 22.880 & 29.669 & 23.011 & 29.533 \\ 
        Bonsai & 31.117 & 24.554 & 31.081 & 25.456 \\
        Room & 30.813 & 23.494 & 30.785 & 23.877 \\ 
        Flowers & 19.483 & 29.518 & 19.476 & 29.272 \\ 
        Treehill & 22.407 & 22.477 & 22.433 & 22.121 \\ 
        Garden & 25.254 & 28.582 & 25.225 & 28.179 \\ 
        Stump & 23.908 & 29.106 & 23.827 & 29.452 \\ 
        Counter & 28.189 & 20.811 & 28.120 & 20.891 \\ 
        Kitchen & 29.777 & 29.065 & 29.862 & 29.064 \\ \midrule
        Average & 25.981 & 26.364 & 25.980 & 26.427 \\ \bottomrule
    \end{tabular}}
    \caption{PSNR scores for comparisons between the proposed method with and without user-specific keys. Due to space constraints, secret scene names are omitted but can be found in~\cref{tab_1hide1}.}
    \label{tab_sup_withoutkey}
\end{table}
\begin{table}[!t]
    \centering
    \resizebox{\linewidth}{!}{
    \begin{tabular}{c|c|c|c|c}
    \toprule
        \multirow{2}{*}{Ratio}& \multicolumn{2}{c|}{Sequentially} & \multicolumn{2}{c}{Randomly} \\ \cmidrule{2-5}
        ~ & $\text{PSNR}_{cover}\uparrow$ & $\text{PSNR}_{secret}\uparrow$ & $\text{PSNR}_{cover}\uparrow$ & $\text{PSNR}_{secret}\uparrow$ \\ \midrule
        5\% & 25.980 & 26.427 & 25.588 & 26.102 \\ 
        10\% & 25.980 & 26.427 & 25.163 & 25.750 \\ 
        15\% & 25.980 & 26.427 & 24.714 & 25.341 \\ 
        25\% & 25.980 & 26.418 & 23.739 & 24.408 \\ \bottomrule
    \end{tabular}}
    \caption{Robustness analysis under different pruning methods.}
    \label{tab_sup_prune}
\end{table}

To rigorously evaluate the robustness of our proposed decoder architecture, we conducted ablation studies comparing performance with and without user-specific key integration. As shown in~\cref{tab_sup_withoutkey}, the decoder maintains comparable performance levels when incorporating user-specific keys relative to the baseline model without keys. This demonstrates that our key integration mechanism successfully enables multi-scene hiding capabilities and wrong key defense functionality without compromising the decoder's reconstruction quality. These results validate that the additional security features do not introduce performance degradation in the primary decoding task.

\subsection{Robustness of the Pruning}
We explore the method's behavior under different Gaussian pruning scenarios, including sequential pruning based on opacity values and random pruning. Sequential pruning removes Gaussians in ascending opacity order, targeting lower-opacity elements first, while random pruning stochastically removes a proportion of Gaussians regardless of opacity. Results in~\cref{tab_sup_prune} suggest reasonable resilience under these modifications, with both pruning strategies showing generally acceptable performance. The steganographic information appears to maintain certain levels of stability even with partial Gaussian removal, indicating potential robustness against structural changes.

\subsection{More Visualization Results}\label{sec_sup_morevis}
\begin{figure*}[t]
\centering
\small
\includegraphics[width=1.0\linewidth]{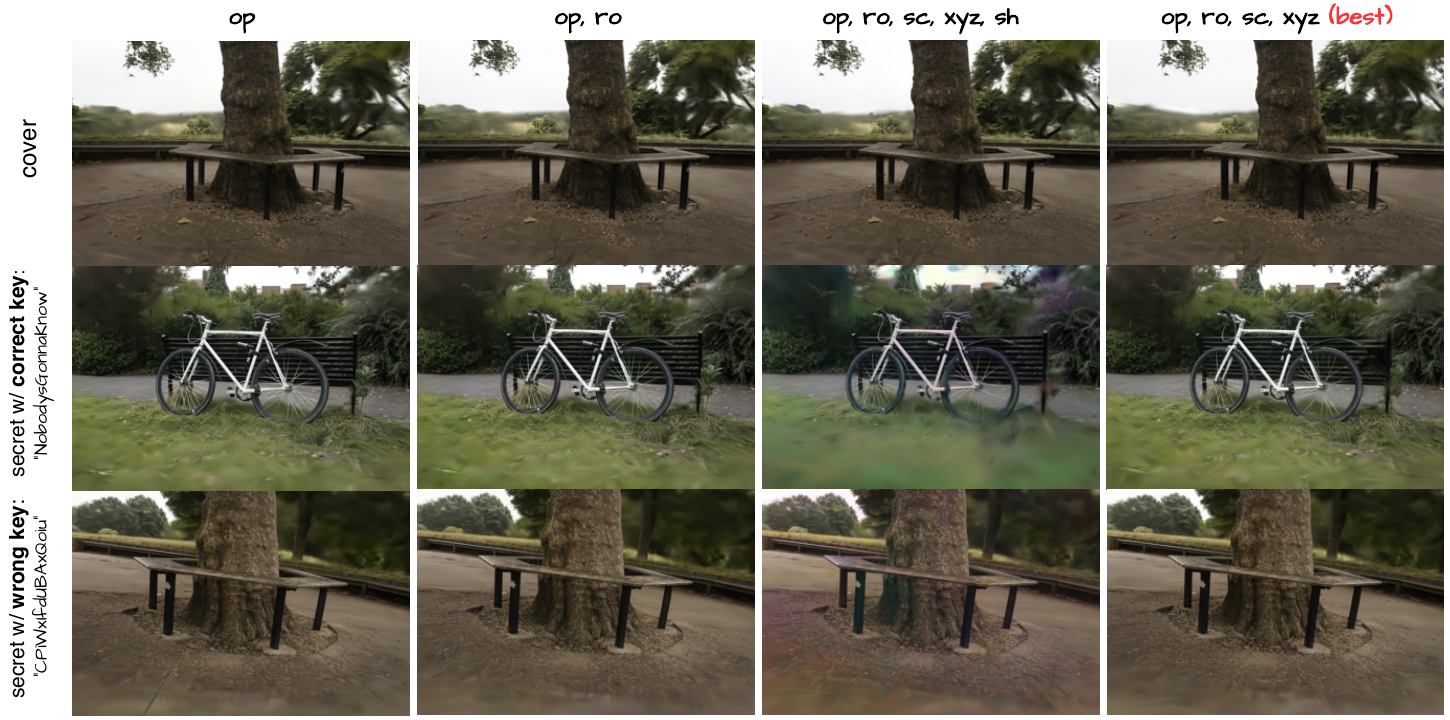}
\includegraphics[width=1.0\linewidth]{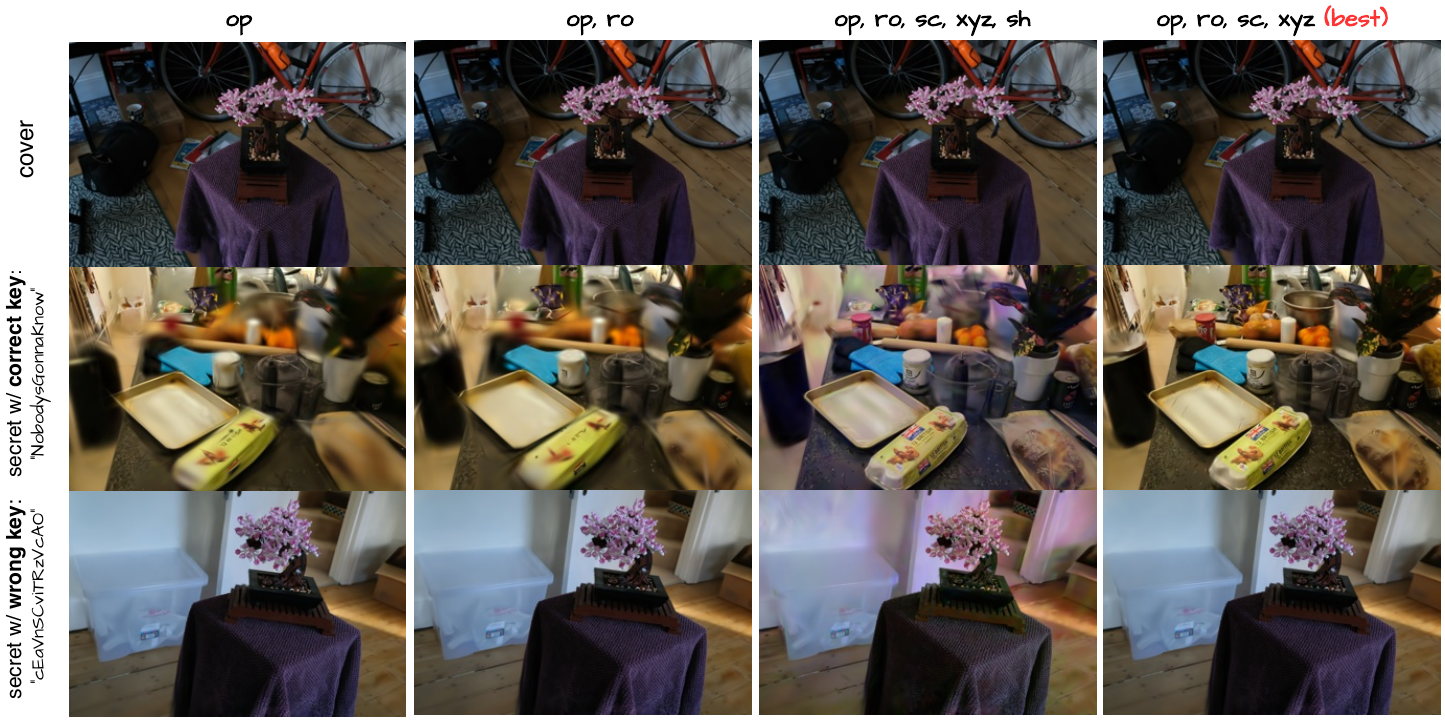}
\caption{More visualizations of the proposed method on hiding single secert across different feature combinations using correct and incorrect keys, which show cover recovery, secret recovery (correct key) and security preservation (incorrect key). Notation follows~\cref{fig_1hid1vis}.}
\label{fig_sup_1hid1vis_single}
\end{figure*}

\begin{figure*}[t]
\centering
\small
\includegraphics[width=1.0\linewidth]{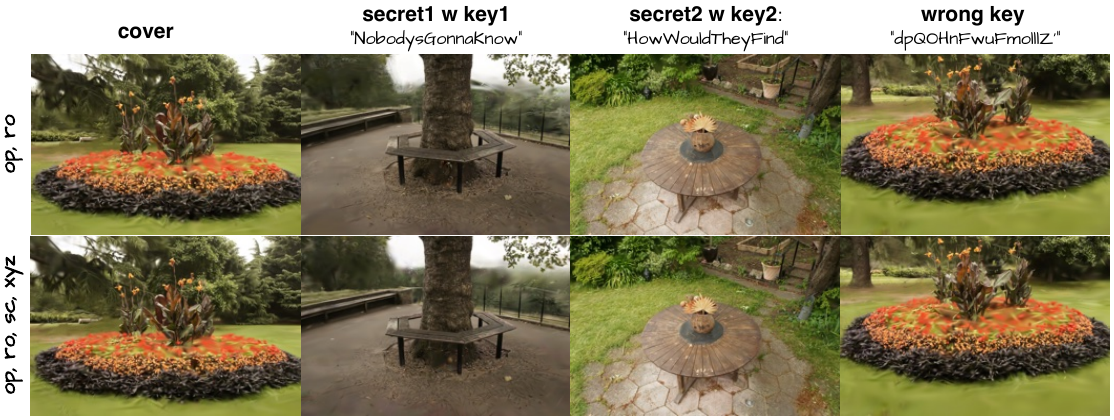}
\includegraphics[width=1.0\linewidth]{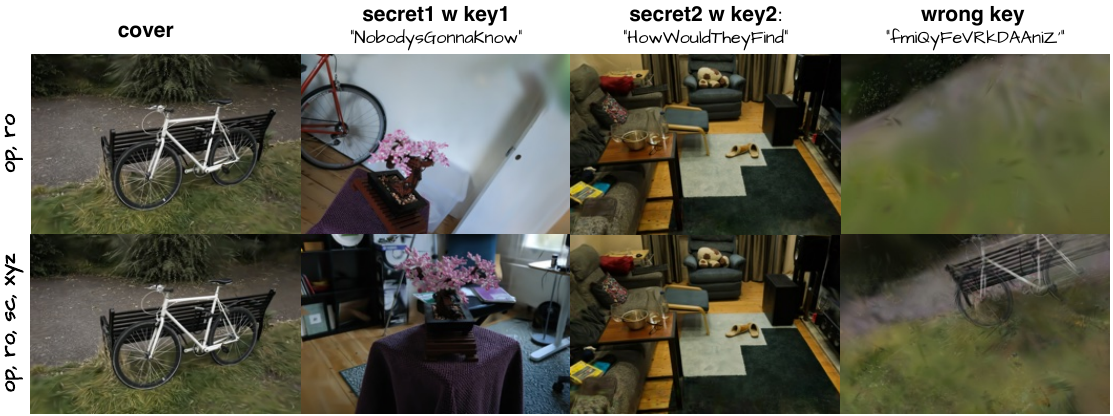}
\caption{More visualizations of the proposed method on hiding multiple secrets across different feature combinations using correct and incorrect keys, which show cover recovery, secret recovery (correct key) and security preservation (incorrect key). Notation follows~\cref{fig_1hid1vis}.}
\label{fig_sup_1hid1vis_multi}
\end{figure*}

Comprehensive visualization results are provided in~\cref{fig_sup_1hid1vis_single,fig_sup_1hid1vis_multi}, offering deeper qualitative insights into the method's performance across diverse scene types. Both single and multiple secret hiding scenarios demonstrate exceptional visual fidelity, with the correctly-keyed reconstructions preserving fine geometric details, texture consistency, and color accuracy. When accessed with incorrect keys, the framework demonstrates robust security properties: unauthorized users receive only the cover-scene-like visualization with no discernible traces of the embedded secrets, as evidenced by the absence of visual artifacts or structural inconsistencies that might otherwise suggest hidden content. These visualizations complement the quantitative metrics by confirming the method's practical effectiveness in maintaining the visual-perceptual balance between high-quality secret reconstruction and security.

\end{document}